\documentclass[12pt,showpacs,amssymb,superscriptaddress]{revtex4}
\usepackage{mathrsfs}

\usepackage{amssymb,amsmath}
\usepackage{amsfonts}

\begin{document}
\title{A complete hybrid quantization in inhomogeneous cosmology}

\author{Mikel Fern\'andez-M\'endez}
\email{m.fernandez.m@csic.es}
\author{Guillermo A. Mena Marug\'an}
\email{mena@iem.cfmac.csic.es}
\author{Javier Olmedo}
\email{olmedo@iem.cfmac.csic.es}
\affiliation{Instituto de Estructura de la Materia, IEM-CSIC, Serrano 121, 28006 Madrid, Spain}


\begin{abstract}
A complete quantization of a homogeneous and isotropic spacetime with closed
spatial sections coupled to a massive scalar field is provided, within the
framework of Loop Quantum Cosmology. We  identify solutions with their initial
data on the minimum volume section, and from this we construct the
physical Hilbert space. Moreover, a perturbative study allows us to introduce
small inhomogeneities. After gauge fixing, the inhomogeneous part of the system is reduced to a linear
field theory.  We then adopt a standard Fock representation to quantize
these degrees of freedom. For the considered case of compact spatial topology,
the requirements of: i) invariance under the spatial  isometries, and ii)
unitary implementation of the quantum dynamics, pick up a unique Fock
representation and a particular set of canonical fields (up to unitary
equivalence).
\end{abstract}

\maketitle

\section{Introduction}

Loop Quantum Cosmology (LQC) is one of the most promising approaches to the quantization of cosmological spacetimes
\cite{lqc}. In particular, its application to homogeneous and isotropic spacetimes
(or even to anisotropic ones \cite{bianchi}) has shown that
the classical singularity is always replaced with a quantum
bounce.

In inflationary scenarios \cite{liddle}, for instance in the presence of a massive
scalar field, a careful analysis assuming some associated effective dynamics
proves the overwhelming probability that the inflaton produces the necessary
amount of e-foldings \cite{as_slow}, solving the fine tuning problem inherent to General
Relativity. But such an effective dynamics does not arise from any genuine
quantization within LQC; it is rather adopted as a reasonable hypothesis. On the other hand, if one introduces small
inhomogeneities --coming from the vacuum fluctuations of the inflaton--, a
valid description for them should be provided by a standard Fock
representation, at least in certain regimes. Nevertheless, the infinite number of possible inequivalent Fock representations
leads to an uncontrollable ambiguity in the physical predictions, unless one can find a
privileged Fock quantization.

In this work, we consider a closed Friedmann-Robertson-Walker (FRW) spacetime coupled to a massive
scalar field. We provide a complete polymeric quantization of this model,
where each solution to the Hamiltonian constraint is characterized by its data
on the minimum volume section. The completion of the space of these data with respect to a suitable
inner product provides the physical Hilbert space. Moreover, the small
inhomogeneities around homogeneous solutions, after a suitable gauge fixing, are
described by a (linear) scalar field theory. Finally, we determine a
unique $SO(4)$-invariant Fock quantization for them by requiring a unitary
implementation of the dynamics.

\section{Homogeneous system}

The homogeneous and isotropic system has two global (canonical pairs of) degrees of freedom. One is
the pair $(\phi,p_\phi)$ for the matter content. Besides, the geometry is
described by a densitized triad, which essentially reduces to $p$ (the square of
the scale factor), and an Ashtekar-Barbero $su(2)$-connection $c$. The definition of these variables takes into account
the $S^3$ topology, and uses a fiducial volume $l^3_0=2\pi^2$ (see
\cite{apsv}). The Poisson bracket is $\{c,p\}=8\pi G \gamma/3$, with $G$ being the
Newton constant and $\gamma$ the Immirzi parameter.

We carry out a polymeric quantization of the geometry degrees of freedom. The
basic variables in LQC are fluxes of densitized triads through surfaces
enclosed by four geodesic edges, which are determined essentially by $p$, and holonomies
of the connection along integral curves of the fiducial triads, with fiducial
length $\bar{\mu}l_0$. Specifically, we adopt the {\it improved dynamics}
scheme, in which $\bar{\mu}=\sqrt{\Delta/p}$, with $\Delta$ equal to the minimum nonzero eigenvalue allowed for the area
in Loop Quantum Gravity \cite{mmo}. Then, the
gravitational part of the kinematical Hilbert space is
$\mathcal{H}^{\rm grav}_{\rm kin}=L^2(\mathbb{R}_{\rm Bohr},d\mu_{\rm Bohr})$,
where $\mathbb{R}_{\rm Bohr}$ is the Bohr compactification of the real line, and
$d\mu_{\rm Bohr}$ is the natural Haar measure associated with it. In this Hilbert
space one can find a basis of normalizable states in which the action of the matrix elements
of the holonomies is given by ${\hat N}_{\bar{\mu}}|v\rangle=|v+1\rangle$,
whereas, for the triad, $\hat p|v\rangle={\rm sgn}(v)(2\pi \gamma G \hbar\sqrt{\Delta}
|v|)^{2/3}|v\rangle$. Here, $\hbar$ is the Planck constant. For the scalar field, on the other hand,
we employ a standard Schr\"odinger representation. Adopting
a suitable factor ordering, we obtain the following operator representation for the Hamiltonian constraint:
\begin{equation}\label{eq:hom_quantum_const}
\hat{C}_0=\widehat{\left[\frac{1}{V}\right]}^{1/2}\!\!\bigg[8\pi G \big(\hat{p}_\phi^2+m^2\hat{V}^2\hat{\phi}^2\big)-
\frac{6}{\gamma^2}\bigg\{\widehat{\Omega}^2+(1+\gamma^2)l_0^2
\hat{V}^{4/3}-\frac{\hat{V}^2}{\Delta}\sin^2\hat{\bar{\mu}}l_0\bigg\}\bigg]\widehat{\left[\frac{1}{V}\right]}^{1/2}\!\!\!\!\!,
\nonumber\end{equation}
where $\widehat{\left[1/V\right]}$ is the regularized inverse of the volume operator $\hat{V}$ (see \cite{apsv,mmo}), and
\begin{eqnarray}
\label{eq:quantumomega}  &\widehat{\Omega} &=
\frac{1}{4i\sqrt{\Delta}}\hat{V}^{1/2}
\Big[\widehat{{\rm sgn}(v)}
\left(e^{-i\frac{\hat{\bar{\mu}}l_0}{2}}\hat{N}_{2\bar{\mu}}e^{-i\frac{\hat{\bar{\mu}}l_0}{2}}-
e^{i\frac{\hat{\bar{\mu}}l_0}{2}}\hat{N}_{-2\bar{\mu}}e^{i\frac{\hat{\bar{\mu}}l_0}{2}}\right)
+\\ \nonumber &+&\left(e^{-i\frac{\hat{\bar{\mu}}l_0}{2}}\hat{N}_{2\bar{\mu}}e^{-i\frac{\hat{\bar{\mu}}l_0}{2}}-
e^{i\frac{\hat{\bar{\mu}}l_0}{2}}\hat{N}_{-2\bar{\mu}}e^{i\frac{\hat{\bar{\mu}}l_0}{2}}\right)\widehat{{\rm sgn}(v)}\Big]\hat{V}^{1/2}.
\end{eqnarray}

The constraint is a second order difference operator that only relates states
with support on semilattices of the type
$\mathcal{L}_{\varepsilon}^{\pm}=\{v=\pm(\varepsilon+4n),n\in\mathbb{N}\}$,
where $\varepsilon\in(0,4]$ is a continuous parameter labeling each subspace,
which can be interpreted as a superselection sector. Moreover, on any solution $\Psi$ to
the constraint, the value at all the volumes $v=\varepsilon+4n$ with $n>0$ is totally determined by the
value on the minimum volume section, namely $\Psi(\phi,v=\varepsilon)$. The physical
Hilbert space can thus be identified with the completion of the functional space of initial data at
$v=\varepsilon$ with respect to an inner product which is uniquely determined by the requirement of
self-adjointness on a complete set of observables. Hence, the physical Hilbert
space of this homogeneous system can be taken~as
\begin{equation}
\mathcal{H}^{\rm phy} = L^2(\mathbb{R},d\phi).
\end{equation}

\section{Inhomogeneous model and gauge fixing}

Small inhomogeneities may arise from vacuum fluctuations
of the inflaton. With this motivation, we introduce inhomogeneities in our model and
adopt a perturbative treatment for them. We
will consider scalar perturbations only (this is possible because vector and tensor modes are
dynamically decoupled from the scalar ones at first perturbative order). Around the
homogeneous solutions, we have $N=\sigma (N_0+\delta N)$, $N_a=\sigma^2\delta N_a$,
$h_{ab}=\sigma^2e^{2\alpha}(\Omega_{ab}+\Delta_{ab})$ and
$\Phi=\sigma^{-1}(l_0^{-3/2}\varphi+\delta \varphi)$, for the lapse, shift, spatial metric and scalar field, respectively. Here
$\sigma^2=4\pi G/3l_0^3$, and $e^{\alpha}$ is
the
scale factor \cite{hh}. Besides, the spatial sections are isomorphic to
$S^3$, for which we have a natural basis of spherical harmonics. In terms of it, we get the expansions:
\begin{eqnarray}\nonumber
\delta N&=&l_0^{3/2}N_0\sum_n g_nQ_n,\quad\quad\quad\quad\quad\quad \delta N_a=l_0^{3/2}e^{\alpha}\sum_n
j_nP_a^n,\\ \nonumber   \Delta_{ab}&=&2l_0^{3/2}\sum_n
a_nQ_n\Omega_{ab}+3b_nP_{ab}^n,\quad \delta \varphi=\sum_n f_nQ_n.
\end{eqnarray}
We will consider perturbations to second order in the Hamiltonian, which takes then the form
$H=N_0\big[C_0+\sum_n(C_{2}^n+g_nC_{1}^n)\big]+\sum_n j_n D_{1}^n$,
where $C_0$ is the scalar constraint of the homogeneous system
($\alpha,\pi_\alpha,\varphi,\pi_\varphi$) [with $p$ replaced with $\alpha$ and $\pi_{\alpha}$ being its momentum],
each $C_{2}^n$ is quadratic in the scalar modes $(a_n,b_n,f_n)$ and their
canonically conjugate momenta $(\pi_{a_n},\pi_{b_n},\pi_{f_n})$,
and $C_{1}^n$ and $D_{1}^n$ are the linear contributions of the perturbations to the scalar and the
diffeomorphism constraints (spanned in Fourier modes), respectively.

In our formalism, the perturbations $g_n$ and $j_n$ are nondynamical variables.
They just play the role of Lagrange  multipliers, associated with first-class
constraints. In order to identify
the true physical degrees of freedom and reduce the model, we now introduce gauge fixing conditions.
One possibility is, e.g., the gauge $a_n=0=b_n$. The derived
conditions $\dot{a}_n=0$ and $\dot{b}_{n}=0$, necessary for consistency with the dynamics, fix the functions $j_n$ and $g_n$.
In addition, the constraints $D_{1}^n=C_1^n=0$ allow us to determine
$\pi_{a_n}$ and $\pi_{b_n}$ in terms of $f_n$ and $\pi_{f_n}$.  Besides, in
order to adapt the system to a form in which one can apply the uniqueness results of
\cite{uniq-S3,uniq2-S3} (concerning its Fock quantization), we will make the change of variables $\bar
f_n=e^{\alpha} f_n$ and
$\bar\pi_{f_n}=e^{-\alpha}[\pi_{f_n}-(3\pi_\varphi^2/\pi_\alpha+\pi_\alpha)
f_n]$, which can be easily extended into a canonical transformation (up to the perturbative order under consideration).

\section{Hybrid quantization}

Assuming that the most important quantum geometry effects are those affecting the homogeneous part of the geometry, we now perform
a complete quantization of the system in which we adopt the polymeric quantization explained above for that homogeneous part, while we adhere to a standard quantization not only for the homogeneous component of the scalar field, but for all the inhomogeneities. Remarkably, we can choose a privileged Fock
quantization for the inhomogeneities: there is only one
Fock representation invariant under the spatial isometries that implements the dynamics unitarily \cite{uniq-S3}.
Actually, this criterion fixes not only the representation of the canonical
commutation relations, but also the choice of a canonical pair for our field \cite{uniq2-S3} (at least if one expects to reach regimes where deparametrization is possible).

In our gauge fixed system, those uniqueness results select the Fock representation which is naturally associated with the massless case for the field described by the modes $\bar f_n$ and $\bar\pi_{f_n}$ \cite{fmo}. The resulting kinematical
Hilbert space is the Fock space $\mathcal{F}$ constructed from
the corresponding one-particle Hilbert space. In total, the
kinematical Hilbert space is $\mathcal{H}^{\rm kin}=\mathcal{H}^{\rm
kin}_{\rm grav}\otimes L^2(\mathbb{R},d\phi)\otimes\mathcal{F}$. With this space as starting point, we can construct a quantum Hamiltonian constraint and look for
its solutions. We expect this constraint to impose a second order difference equation with solutions totally determined by their values on the minimum volume section (at least in the studied perturbative order). Identifying then solutions
with initial data, the latter can be completed with a suitable inner product
to obtain the physical Hilbert space (see
e.g. \cite{gmm} and the previous section).

\section{Conclusions}

We have first analyzed a homogeneous, massive scalar field propagating in
a closed FRW spacetime and quantized the system to completion, combining
both Schr\"odinger and polymeric representations. The physical Hilbert space has been
constructed out of the functional space defined on the minimum volume section,
which provides the initial data for the solutions of the scalar constraint.
Along similar lines, we can also obtain
the physical Hilbert space when small
inhomogeneities are included. These are introduced in the system as perturbations.
In the process followed to arrive to this space of physical states, we have carried out a partial gauge fixing,
removing two of the three fieldlike degrees of freedom, and performed a canonical transformation (consisting in a scaling of the field)
to write the reduced system in a way in which one can appeal to the uniqueness theorems of \cite{uniq-S3,uniq2-S3}.
Finally, we have adopted a hybrid quantization for this reduced system, combining loop techniques for
the homogeneous sector of the geometry with a standard Fock quantization of the inhomogeneities
\cite{fmo}.

We acknowledge the MICINN/MINECO Projects FIS2008-06078-C03-03 and FIS2011-30145-C03-02, and
the Consolider-Ingenio Program CPAN CSD2007-00042.



\begin{thebibliography}{M}


\bibitem[Bojowald(2008)]{lqc} Bojowald M 2008 {\it Living Rev. Relativity} {\bf 11}, 4

\bibitem[Ashtekar(2009)]{bianchi} Ashtekar A and Wilson-Ewing E
2009 {\it Phys. Rev.} D {\bf 79}, 083535; Mart\'in-Benito M, Mena Marug\'an G A
and Paw{\l}owski T 2008 {\it Phys. Rev.} D {\bf 78}, 064008

\bibitem[Liddle(2000)]{liddle} Liddle A R and Lyth D H 2000 {\it Inflation
and Large-Scale Structure} (Cambridge University Press)

\bibitem[Ashtekar and Sloan(2010)]{as_slow} Ashtekar A and Sloan D 2010 {\it Phys. Lett.} B {\bf 694},
108

\bibitem[Ashtekar et al(2007)]{apsv} Ashtekar A, Paw{\l}owski T, Singh P and Vandersloot K 2007 {\it
Phys. Rev.} D {\bf 75}, 024035

\bibitem[Martin-Benito et al(2009)]{mmo}
Mart\'in-Benito M, Mena Marug\'an G A and Olmedo J 2009 {\it Phys. Rev.} D {\bf
80}, 104015

\bibitem[Halliwell and Hawking(1985)]{hh} Halliwell J J and Hawking W 1985 {\it Phys. Rev.} D {\bf 31}, 8

\bibitem[Cortez et al(2010)]{uniq-S3} Cortez J, Mena Marug\'an G A and Velhinho J M 2010 {\it Phys.
Rev.} D {\bf 81}, 044037

\bibitem[Cortez et al(2010)]{uniq2-S3} Cortez J, Mena Marug\'an G A, Olmedo J and Velhinho J M 2010
{\it JCAP} {\bf 10}, 030

\bibitem[Fern\'andez-M\'endez et al(2012)]{fmo} Fern\'andez-M\'endez M, Mena
Marug\'an G A, Olmedo J and Velhinho J M 2012 {\it Phys. Rev.} D {\bf 85}, 103525

\bibitem[Garay et al(2010)]{gmm} Garay L J, Mart\'in-Benito M and Mena Marug\'an G A 2010 {\it
Phys. Rev.} D {\bf 82}, 044048


\end{thebibliography}
\end{document}